\documentclass[twocolumn,showpacs,prl,aps]{revtex4}
\usepackage{amssymb}
\usepackage{amsmath}
\usepackage{amsfonts}
\usepackage{graphics}
\usepackage{epic}
\usepackage{eepic}
\usepackage{color}
\usepackage{epsfig}

\begin{document}

\def\ra{\rangle}
\def\la{\langle}
\def\bege{\begin{equation}}
\def\ende{\end{equation}}
\def\begarr{\begin{eqnarray}}
\def\endarr{\end{eqnarray}}
\def\ha{{\hat a}}
\def\hb{{\hat b}}
\def\hu{{\hat u}}
\def\hv{{\hat v}}
\def\hc{{\hat c}}
\def\hd{{\hat d}}
\def\no{\noindent}\def\non{\nonumber}
\def\hi{\hangindent=45pt}
\def\v{\vskip 12pt}

\newcommand{\bra}[1]{\left\langle #1 \right\vert}
\newcommand{\ket}[1]{\left\vert #1 \right\rangle}

\title{Improving the Fidelity of Optical Zeno Gates via Distillation}

\author{Patrick M. Leung} \email{pmleung@physics.uq.edu.au}
\author{Timothy C. Ralph} 

\affiliation{Centre for Quantum Computer Technology, Department of
Physics,  University of Queensland, Brisbane 4072, Australia}

\date{\today}

\pacs{03.67.Lx, 42.50.-p}

\begin{abstract}
We have modelled the Zeno effect Control-Sign gate of Franson et al
(PRA 70, 062302, 2004) and shown that high two-photon to one-photon
absorption ratios, $\kappa$, are needed for high fidelity free
standing operation. Hence we instead employ this gate for cluster
state fusion, where the requirement for $\kappa$ is less
restrictive. With the help of partially offline one-photon and
two-photon distillations, we can achieve a fusion gate with unity
fidelity but non-unit probability of success. We conclude that for
$\kappa > 2200$, the Zeno fusion gate will out perform the
equivalent linear optics gate.
\end{abstract}

\maketitle


\section{Introduction}

Quantum bits (qubits) based on polarization or spatial degrees of
freedom of optical modes have several advantages: they are easily
manipulated and measured; they exist in a low noise environment and;
they are easily communicated over comparitively long distances.
Recently considerable progress has been made on implementing two
qubit gates in optics using the measurement induced non-linearities
proposed by Knill, Laflamme and Milburn \cite{KNI01}.
Non-deterministic experimental demonstrations have been made
\cite{PIT03, OBR03, GAS04} and theory has found significant ways to
reduce the resource overheads \cite{YOR03,NIE04,HAY04,BRO05}.
Nevertheless, the number of photons and gate operations required to
implement a near deterministic two qubit gate remains high.

A possible solution to this problem is the optical quantum Zeno gate
suggested by Franson et al \cite{ref:Franson}, ~\cite{ref:Jacobs}.
This gate uses passive two-photon absorption to suppress gate
failure events associated with photon bunching at the linear optical
elements, using the quantum Zeno effect \cite{ref:Kaiser}. In
principle a near deterministic, high fidelity control-sign (CZ) gate
can be implemented between a pair of photonic qubits in this way.
However, the slow convergence of the Zeno effect to the ideal
result, with ensuing loss of fidelity, and the effect of single
photon loss raises questions about the practicality of this
approach.

Here we consider a model of the gate that includes the effects of
finite two-photon absorption and non-negligible single photon
absorption. We obtain analytic expressions for the fidelity of the
gate and its probability of success in several scenarios and show
how the inclusion of optical distilling elements~\cite{ref:Thew} can
lead to high fidelity operation under non-ideal conditions for tasks
such as cluster state construction~\cite{NIE04}.

The paper is arranged in the following way. We begin in the next
section by introducing our model in an idealized and then more
realistic setting and obtain results for a free-standing CZ gate. In
section 3 we focus on using the gate as a fusion
element~\cite{BRO05} for the construction of, for example, optical
cluster states. We introduce a distillation protocol that
significantly improves the operation of the gate in this scenario.
In section 4 we summarize and conclude.\\

\section{Model of Zeno CZ Gate}

Franson et al~\cite{ref:Franson} suggested using a pair of optical
fibres weakly evanescently coupled and doped with two-photon
absorbing atoms to implement the gate. As the photons in the two
fibre modes couple the occurence of two photon state components is
suppressed by the presence of the two-photon absorbers via the Zeno
effect. After a length of fibre corresponding to a complete swap of
the two modes a $\pi$ phase difference is produced between the $|11
\rangle$ term and the others. If the fibre modes are then swapped
back by simply crossing them, a CZ gate is achieved.

\begin{figure}[h]
\centerline{\psfig{figure=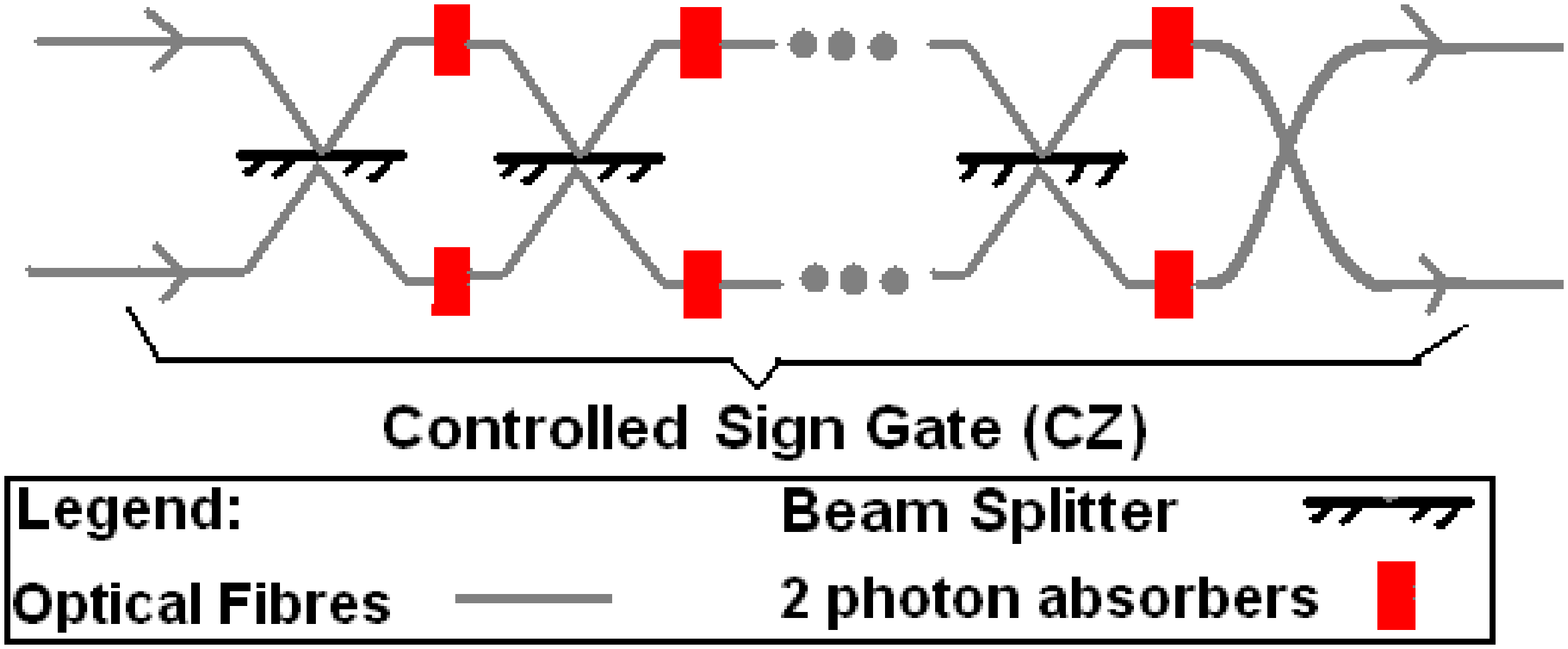,width=2.5cm,angle=90}}
\caption{Construction of our CZ gate.}
\label{fig:OurCsign}
\end{figure}

\begin{figure}[h]
\centerline{\psfig{figure=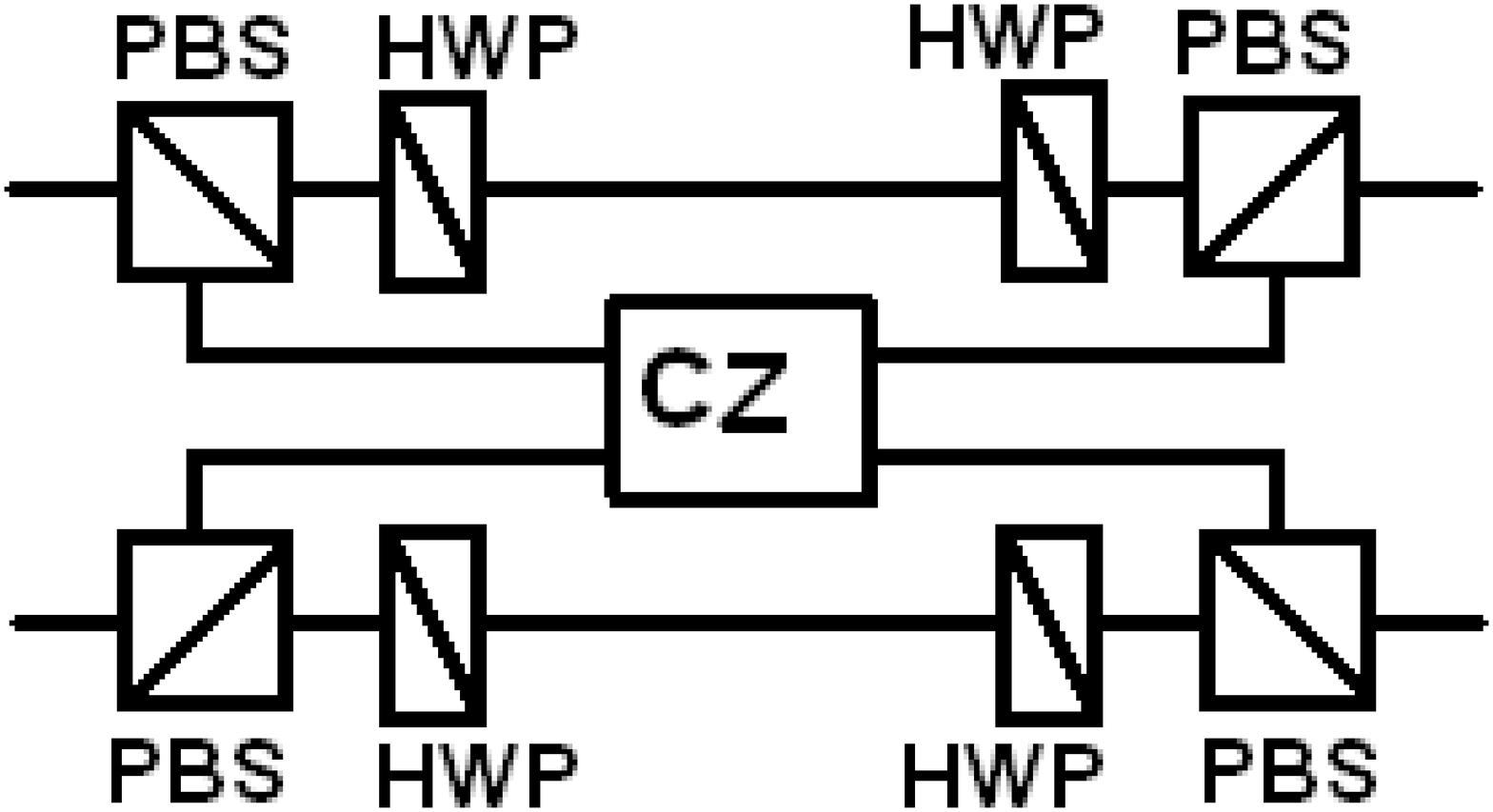,width=2cm,angle=90}} \caption{CZ
gate in dual rail implementation.}\label{fig:CZ}
\end{figure}

We model this system as a succession of $n$ weak beamsplitters
followed by 2-photon absorbers as shown in Fig.~\ref{fig:OurCsign}.
As $n \to \infty$ the model tends to the continuous coupling limit
envisaged for the physical realization. The gate operates on the
single-rail encoding~\cite{LUN02} for which $|0\ra_{L}=|0\ra$ and
$|1\ra_{L}=|1\ra$ with the kets representing photon Fock states.
Fig.~\ref{fig:CZ} shows how the single rail CZ can be converted into
a dual rail CZ with logical encoding $|0\ra_{L}=|H\ra=|10\ra$ and
$|1\ra_{L}=|V\ra=|01\ra$ with $|ij\ra$ a Fock state with $i$ photons
in the horizontal polarization mode and $j$ photons in the
vertical.\\

The general symmetric beam splitter matrix has the form:


\[ e^{i\delta} \left[ \begin{array}{cc}
\cos\theta & \pm i\sin\theta \\
\pm i\sin\theta & \cos\theta \end{array} \right]\]

According to Figure~\ref{fig:OurCsign}, after the first beam
splitter, the four computational photon number states become:
\begin{eqnarray}
|00\ra & \rightarrow & |00\ra \nonumber\\
|01\ra & \rightarrow & e^{i\delta}(\cos\theta|01\ra \pm i\sin\theta|10\ra)\nonumber\\
|10\ra & \rightarrow & e^{i\delta}(\pm i\sin\theta|01\ra +\cos\theta|10\ra)\nonumber\\
|11\ra & \rightarrow & e^{i2\delta}(\cos2\theta|11\ra \pm
\frac{i}{\sqrt{2}}\sin2\theta(|02\ra+|20\ra))
\end{eqnarray}

\subsection{Ideal Two-Photon Absorption}

To illustrate the operation of the gate we first assume ideal two-photon absorbers, i.e. they completely
block the two-photon state components but do not cause any single photon loss. Propagation through the first pair of ideal two-photon absorbers gives the mixed state
\begin{equation}
\rho^{(1)} = P_s^{(1)} | \phi \rangle^{(1)} \langle \phi|^{(1)} + P_f^{(1)} |vac \rangle \langle vac|
\end{equation}
where $|\phi \rangle^{(1)}$ is the evolved two-mode input state obtained for the case of no two-photon absorption event and $|vac \rangle$ is the vacuum state obtained in the case a two-photon absorption event occurs. The individual components of $|\phi \rangle^{(1)}$ transform as
\begin{eqnarray}
|00\ra & \rightarrow & |00\ra \nonumber\\
|01\ra & \rightarrow & e^{i\delta}(\cos\theta|01\ra \pm i\sin\theta|10\ra)\nonumber\\
|10\ra & \rightarrow & e^{i\delta}(\pm i\sin\theta|01\ra +\cos\theta|10\ra)\nonumber\\
|11\ra & \rightarrow & e^{i2\delta}\cos2\theta|11\ra
\label{eqn:complete}
\end{eqnarray}

Notice that, because we are embedded in a dual rail circuit, we can distinguish
between the $|00\ra$ state that corresponds to input state $|HH\ra$
and the $|vac \ra$ state that results from two-photon absorption of
input state $|VV\ra$.

Equation~(\ref{eqn:complete}) describes the transformation of
each unit, hence repeating the procedure $n$ times gives,
\begin{eqnarray}
|00\ra & \rightarrow & |00\ra\nonumber\\
|01\ra & \rightarrow & e^{in\delta}(\cos n\theta|01\ra \pm i\sin n\theta|10\ra)\nonumber\\
|10\ra & \rightarrow & e^{in\delta}(\pm i\sin n\theta|01\ra + \cos n\theta|10\ra)\nonumber\\
|11\ra & \rightarrow & e^{i2n\delta}(\cos2\theta)^n|11\ra
\end{eqnarray}
describing the transformations giving the evolved input state after $n$ units, $|\phi \rangle^{(n)}$.
There are three conditions to satisfy for building a CZ gate. The
first condition is ``~$n\theta=\frac{\pi}{2}$~", so that $|01\ra
\rightarrow e^{i(n\delta\pm\frac{\pi}{2})}|10\ra$ and $|10\ra
\rightarrow e^{i(n\delta\pm\frac{\pi}{2})}|01\ra$. The second
condition is ``$n\delta\pm\frac{\pi}{2}=k\pi$" (equivalently,
$\delta=\frac{\pi}{2n}+\frac{k\pi}{n}$), where $k$ is any integer,
so that $|10\ra \rightarrow e^{ik\pi}|01\ra$ and $|01\ra \rightarrow
e^{ik\pi}|10\ra$ and $|11\ra \rightarrow -(\cos 2\theta)^n|11\ra$.
Phase shifters are needed to correct the sign of the output state of
$|01\ra$ and $|10\ra$ for odd $k$, but here we simply set $k=0$. The
last condition is ``$\cos2\theta > 0$" (i.e. $0< \theta <
\frac{\pi}{4}$), such that a minus sign is induced on $|11\ra$. This
condition is always true because we are using many weak beam
splitters (i.e. $\theta$ is small). Let $\tau = (\cos 2\theta)^n =
(\cos\frac{\pi}{n})^n \ge 0$, then swapping the fibres gives the transformations
\begin{eqnarray}
|00\ra & \rightarrow & |00\ra\nonumber\\
|01\ra & \rightarrow & |01\ra\nonumber\\
|10\ra & \rightarrow & |10\ra\nonumber\\
|11\ra & \rightarrow & -\tau|11\ra
\end{eqnarray}

Clearly, the above is a controlled sign operation with a skew
(quantified as $\tau$) on the probability amplitude of the $|11\ra$
state. If there is some way to herald failure, i.e. two-photon absorption events, then the fidelity of the gate will be $F_h = |\langle T|\phi \rangle^{(n)}|^2$, where $|T \rangle$ is the target state, and the probability of success will be $P_s^{(n)}$. On the other hand if two-photon absorption events are unheralded then the fidelity will be $F_{uh} = F_h P_s^{(n)}$. For simplicity we consider the equally weighted superposition input state
$\frac{1}{2}(|00\ra+|01\ra+|10\ra+|11\ra)$. The corresponding $|\phi \rangle^{(n)}$ after the Zeno-CZ gate is
$\frac{1}{2}(|00\ra+|01\ra+|10\ra-\tau|11\ra)$, to be compared with the target state $|T \rangle = \frac{1}{2}(|00\ra+|01\ra+|10\ra- |11\ra)$.  The heralded fidelity and probability of success
are then $F_h =\frac{(3+\tau)^2}{4(3+\tau^2)}$ and $P_s=\frac{3+\tau^2}{4}$
respectively. As $n$ becomes very
large and hence tends to the continuous limit, $\tau$ tends to one, and so both $F_h$ and $P_s$ approach one.

\subsection{Incomplete Two-Photon Absorption with Single Photon Loss}

The previous analysis is clearly unrealistic as it assumes
infinitely strong two-photon absorption but negligible single photon
absorption. We now include the effect of finite two-photon
absorption and non-negligible single photon loss. Let
$\gamma_{1}=\exp(\frac{-\lambda}{n\kappa})$ and
$\gamma_{2}=\exp(\frac{-\lambda}{n})$ be the probability of single
photon and two-photon transmission respectively for one absorber.
Here the parameter $\lambda=\chi L$, where $L$ is the length of the
absorber and $\chi$ is the corresponding proportionality constant
related to the absorption cross section. Furthermore, $\kappa$
specifies the relative strength of the two transmissions and relates
them by $\gamma_{2}=\gamma_{1}^{\kappa}$. Now each unit of weak beam
splitter and absorbers does the following transformation on the
computational states of $|\phi \rangle$

\begin{eqnarray}
|00\ra & \rightarrow & |00\ra \nonumber\\
|01\ra & \rightarrow & e^{i\delta}\sqrt{\gamma_{1}}(\cos\theta|01\ra \pm i\sin\theta|10\ra)\nonumber\\
|10\ra & \rightarrow & e^{i\delta}\sqrt{\gamma_{1}}(\pm i\sin\theta|01\ra +\cos\theta|10\ra)\nonumber\\
|11\ra & \rightarrow & e^{i2\delta}\gamma_{1}\bigl(\cos2\theta|11\ra
\pm \frac{i\sqrt{\gamma_{2}}\sin2\theta}{\sqrt{2}}(|02\ra+|20\ra)\bigr)\nonumber\\
|02\ra & \rightarrow & e^{i2\delta}\gamma_{1}(\frac{\pm
i\sin2\theta}{\sqrt{2}}|11\ra +
\sqrt{\gamma_{2}}(\cos^2\theta|02\ra-\sin^2\theta|20\ra))\nonumber\\
|20\ra & \rightarrow & e^{i2\delta}\gamma_{1}(\frac{\pm
i\sin2\theta}{\sqrt{2}}|11\ra -
\sqrt{\gamma_{2}}(\sin^2\theta|02\ra-\cos^2\theta|20\ra))\nonumber\\
\end{eqnarray}

Repeating the procedure $n$ times with the aforementioned conditions on $\theta$
gives the following
\begin{eqnarray}
|00\ra & \rightarrow & |00\ra\nonumber\\
|01\ra & \rightarrow & \gamma_{1}^{n/2}|01\ra\nonumber\\
|10\ra & \rightarrow & \gamma_{1}^{n/2}|10\ra\nonumber\\
|11\ra & \rightarrow & -\gamma_{1}^{n}\tau|11\ra + f(|02 \rangle,
|20 \rangle) \label{eqn:incomplete}
\end{eqnarray}
where the new expression for $\tau$ is given by:
\begin{eqnarray}
\tau_{n,\lambda} & = & \frac{2^{-\frac{3}{2}-n}}{d}\bigl( (g+\frac{d}{\sqrt{2}})^{n}(\sqrt{2}d-h) \nonumber\\
& & + (g-\frac{d}{\sqrt{2}})^{n}(\sqrt{2}d+h)\bigr)\nonumber\\
d_{n,\lambda} & = & \sqrt{(1+\cos\frac{2\pi}{n})(1+\gamma_{2})+2\sqrt{\gamma_{2}}(\cos(\frac{2\pi}{n})-3)}\nonumber\\
g_{n,\lambda} & = & (\cos\frac{\pi}{n})(\sqrt{\gamma_{2}}+1)\nonumber\\
h_{n,\lambda} & = & 2(\cos\frac{\pi}{n})(\sqrt{\gamma_{2}}-1)
\end{eqnarray}
and we have suppressed the explicit form of the $|02 \rangle, |20
\rangle$ state components as they lie outside the computational
basis and so do not explicitly contribute to the fidelity. These
expressions can be used to calculate the unheralded fidelity, the
heralded fidelity and probability of success. Our numerical
evaluations are all carried out in the (near) continuous limit of
large $n$.

\subsection{Free Standing Gate}

\begin{figure}[h]
\centerline{\psfig{figure=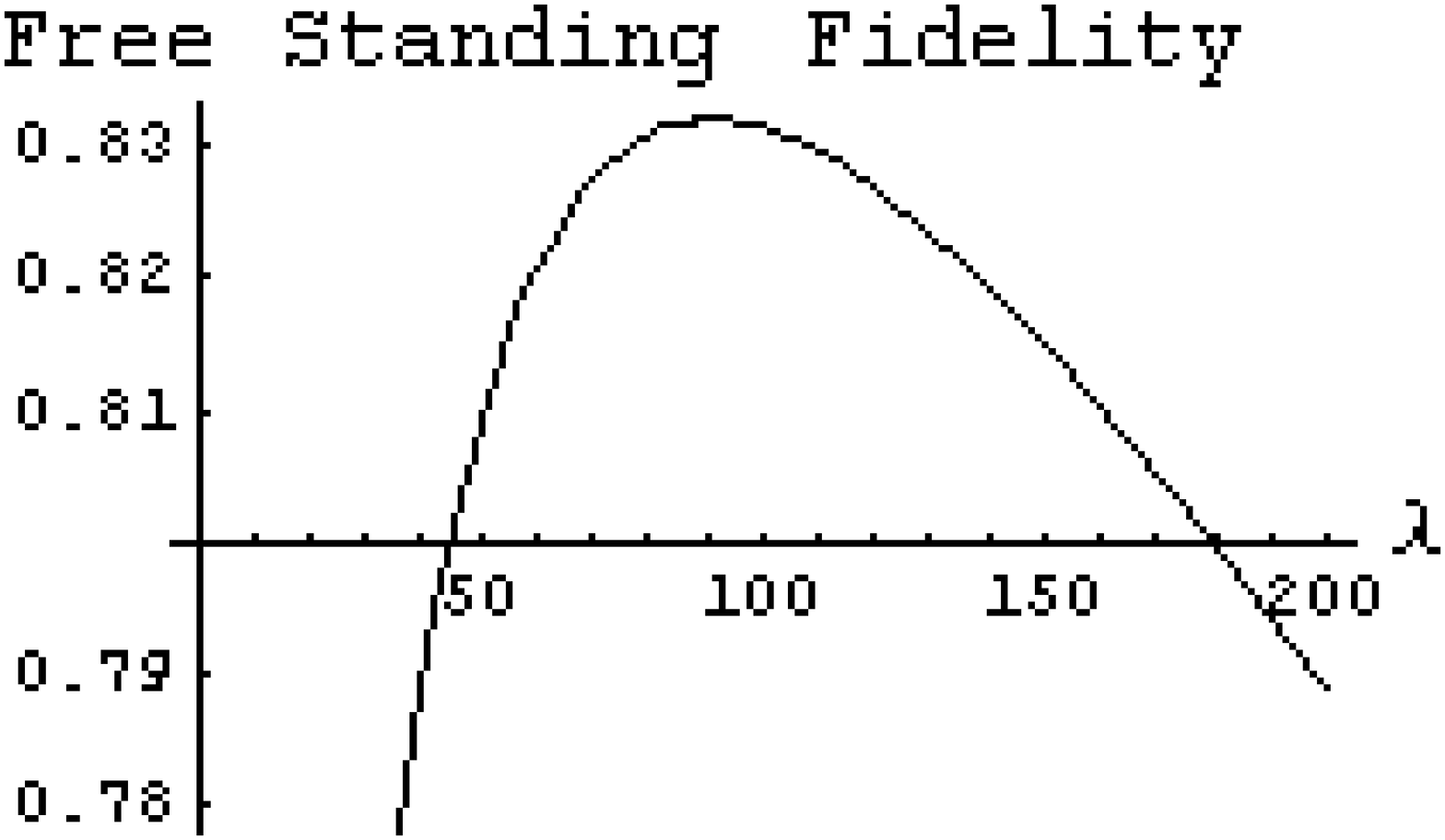,width=4cm,angle=90}}
\caption{Unheralded fidelity versus $\lambda$ for the CZ gate shown
in Fig.\ref{fig:CZ}. Here $\kappa=1000$.}
\label{fig:fidelity_vs_lambda}
\end{figure}

\begin{figure}[h]
\centerline{\psfig{figure=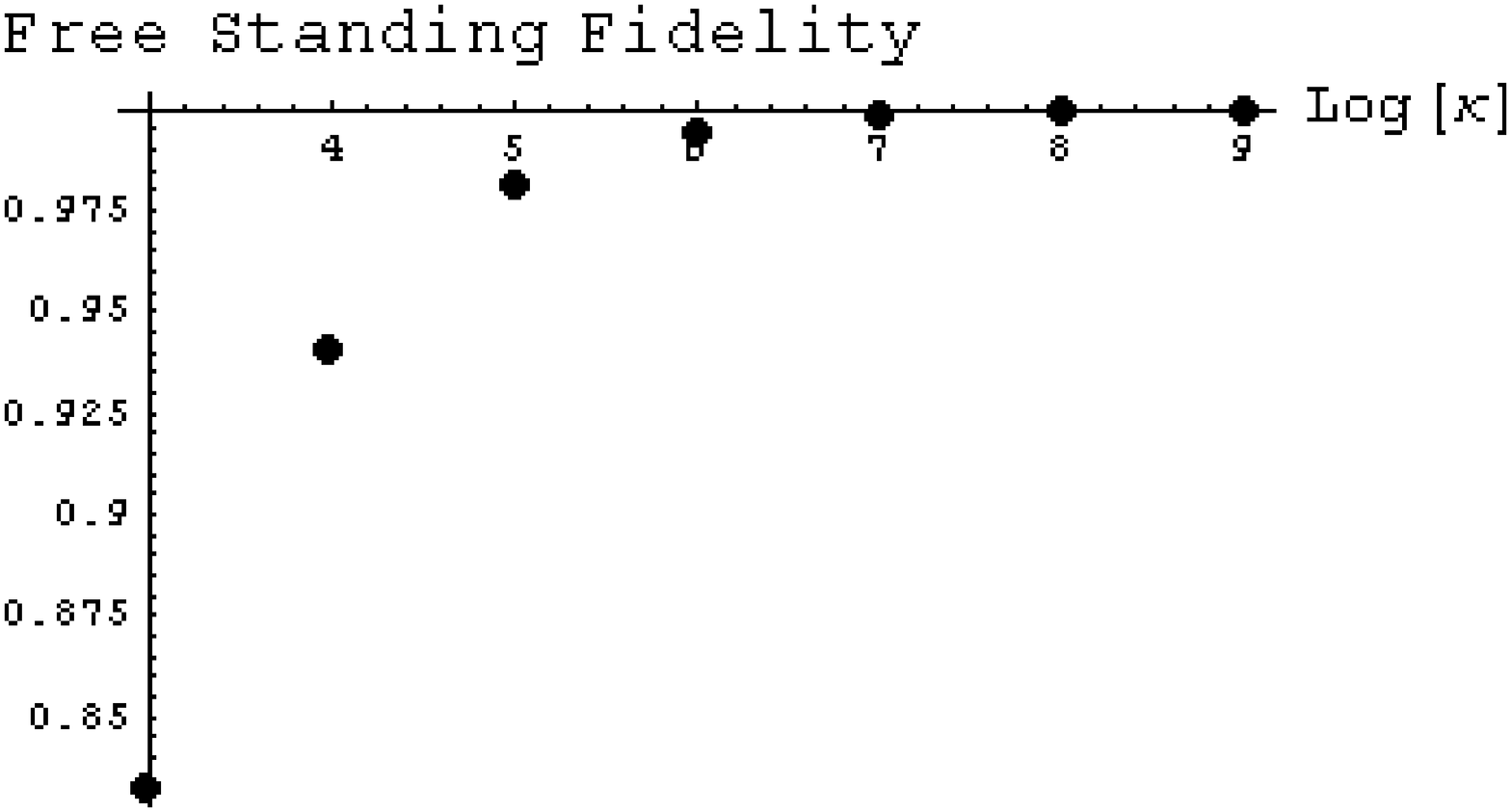,width=3.5cm,angle=90}}
\caption{Unheralded fidelity versus Log($\kappa$) (in base 10) for
the CZ gate shown in Fig.\ref{fig:CZ}, where we have used the
optimal values of $\lambda$.}\label{fig:fidelity_vs_kappa}
\end{figure}

For a free-standing gate, as depicted in Fig.\ref{fig:CZ}, gate
failure events are not heralded, thus the unheralded fidelity is
appropriate to consider. The fidelity is a function of $\lambda$. As
the length of the interaction region is increased ($\lambda$
increased) the effective strength of the two-photon absorption is
increased leading to an improvement in the heralded fidelity, $F_h$.
However, at the same time, the level of single photon absorption is
also increasing with the length, acting to decrease the probability
of success, $P_s$. As the unheralded fidelity is $F_{uh} = F_h P_s$,
there is a trade-off between these two effects leading to an optimum
value for $\lambda$ for sufficiently large $\kappa$. An example of
the dependence is shown in Fig.\ref{fig:fidelity_vs_lambda}. The
fidelity is plotted as a function of $\kappa$ with $\lambda$
optimized for each point in Fig.\ref{fig:fidelity_vs_kappa}. For
large ratios of two-photon absorption to single-photon absorption,
$\kappa$, we tend to the ideal case of unit fidelity. However, the
conditions required are demanding with absorption ratios of a
million to one required for $F_{uh} > 0.99$ and 100 million to one
for $F_{uh} > 0.999$. Recent estimates suggest $\kappa$'s of ten
thousand to one may be achievable~\cite{ref:Franson3}, well short of
these numbers. In the following we will consider a different
scenario in which the gate can be usefully employed with less
stringent conditions on $\kappa$.\\

\section{Zeno Fusion Gate}

We have seen that the requirements on high fidelity operation for
the free-standing gate are quite extreme. We now consider an
alternate scenario in which probability of success is traded-off
against fidelity by heralding failure events through direct
detection. In particular we consider using the Zeno gate to
implement the fusion technique \cite{BRO05}. Fusion can be used to
efficiently construct cluster states \cite{ref:RAU01}, or re-encode
parity states \cite{ref:Ralph}. We will specifically consider
cluster state construction here. Essentially, the gate is used to
make a Bell measurement on a pair of qubits, as depicted in
Fig.\ref{fig:fusion}. One of the qubits comes from the cluster we
are constructing, whilst the other comes from a resource cluster
state, in a known logical state. The Bell measurement has the effect
of ``fusing" the resource state onto the existing state. By careful
choice of the resource state, large 2-dimensional cluster states,
suitable for quantum computation, can be
constructed~\cite{ref:Dawson}. Because the Bell measurement ends
with the direct detection of the qubits, the loss of one or both of
the photons, or the bunching of two photons in a single qubit mode
can immediately be identified in the detection record, and hence
failure events will be heralded. Effectively we will postselect the
density operator $\rho = P_s | \phi' \rangle \langle \phi'| +
\rho_r$, where $|\phi' \rangle$ is the component of the output state
which remains in the computational basis and $\rho_r$ are all the
components that do not. The measurement record then allows us to
herald the first term of the density operator as successful
operation, with fidelity $F_h = |\langle T|\phi' \rangle|^2$ and
probability of success of $P_s$, and the second term as failure. We
now consider techniques for improving the heralded fidelity of the
gate and then evaluate its performance as a fusion gate.

\subsection{Single Photon Distillation}

\begin{figure}[h]
\centerline{\psfig{figure=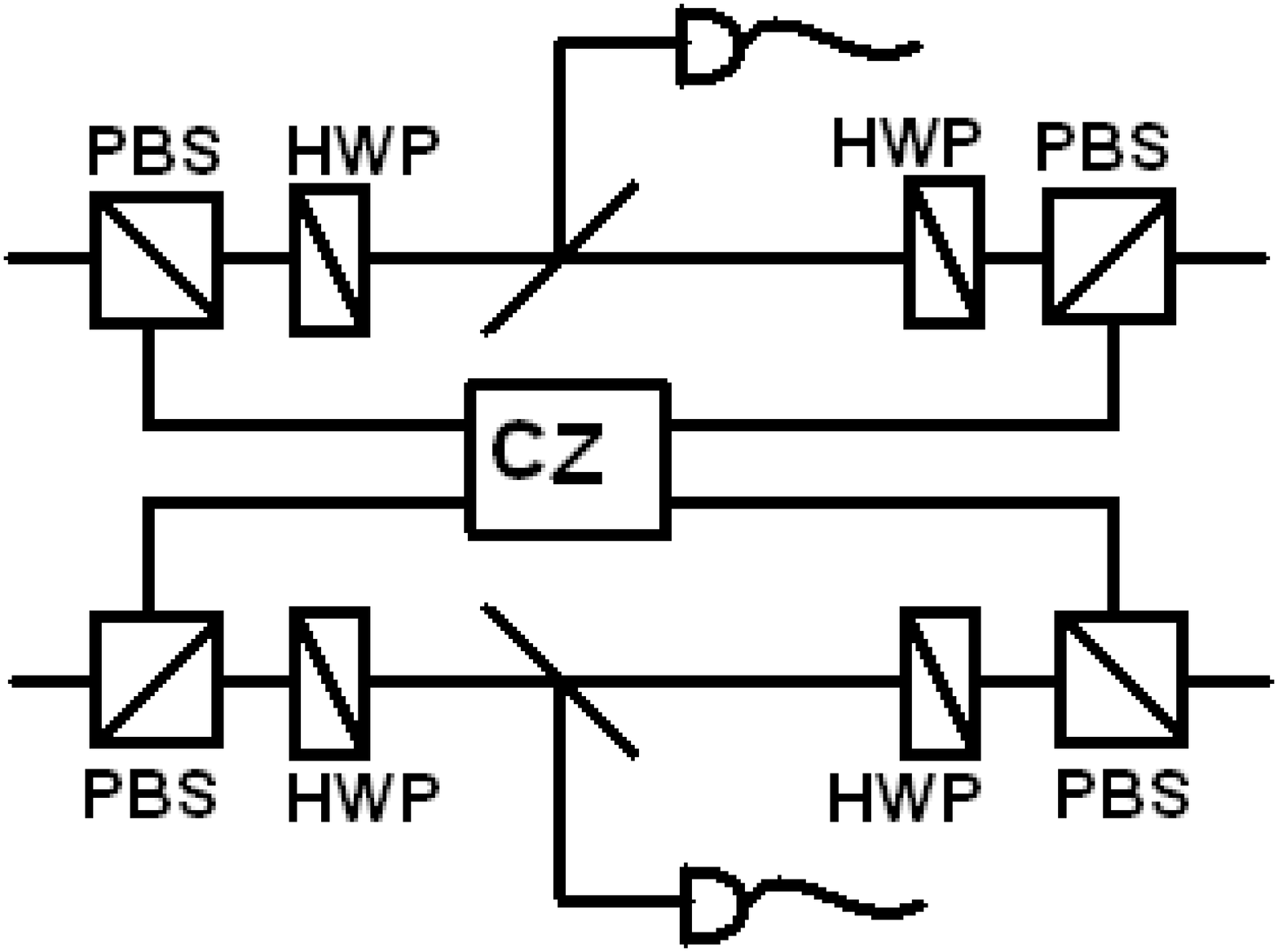,width=2.5cm,angle=90}}
\vspace{0.2cm}\caption{CZ gate in dual rail implementation with beam
splitter distillation to improve average
fidelity.}\label{fig:cz_distill}
\end{figure}

 From equation~(\ref{eqn:incomplete}),
we can see that $\gamma_{1}<1$ lowers the probability amplitude of
the four computational states unevenly as previously discussed by Jacobs et al~\cite{ref:Jacobs}. By
distilling the states with beam splitters and detectors \cite{ref:Thew}  (see
Figure~\ref{fig:cz_distill}), where each beam splitter has a
transmission coefficient equal to $\gamma_{1}^{n/2}$, the four
computational states of $|\phi' \rangle$ become:
\begin{eqnarray}
|00\ra & \rightarrow & \gamma_{1}^{n}|00\ra\nonumber\\
|01\ra & \rightarrow & \gamma_{1}^{n}|01\ra\nonumber\\
|10\ra & \rightarrow & \gamma_{1}^{n}|10\ra\nonumber\\
|11\ra & \rightarrow & -\gamma_{1}^{n}\tau|11\ra
\end{eqnarray}

The distillation is successful when the (ideal) detectors measure no
photon. The fidelity and probability of success of this scheme are
$F_h=\frac{(3+\tau)^2}{4(3+\tau^2)}$ and
$P_s=\gamma_{1}^{2n}(\frac{3+\tau^2}{4}) =
e^{-2\lambda/\kappa}(\frac{3+\tau^2}{4})$ respectively.

For $\lambda$ tends to infinity $F_h \to 1$, however at the same time $P_s \to 0$. In order to achieve unit
fidelity \emph{independent} of $\lambda$, we now apply two-photon
distillation.

\subsection{Two-Photon Distillation}

\begin{figure}[h]
\centerline{\psfig{figure=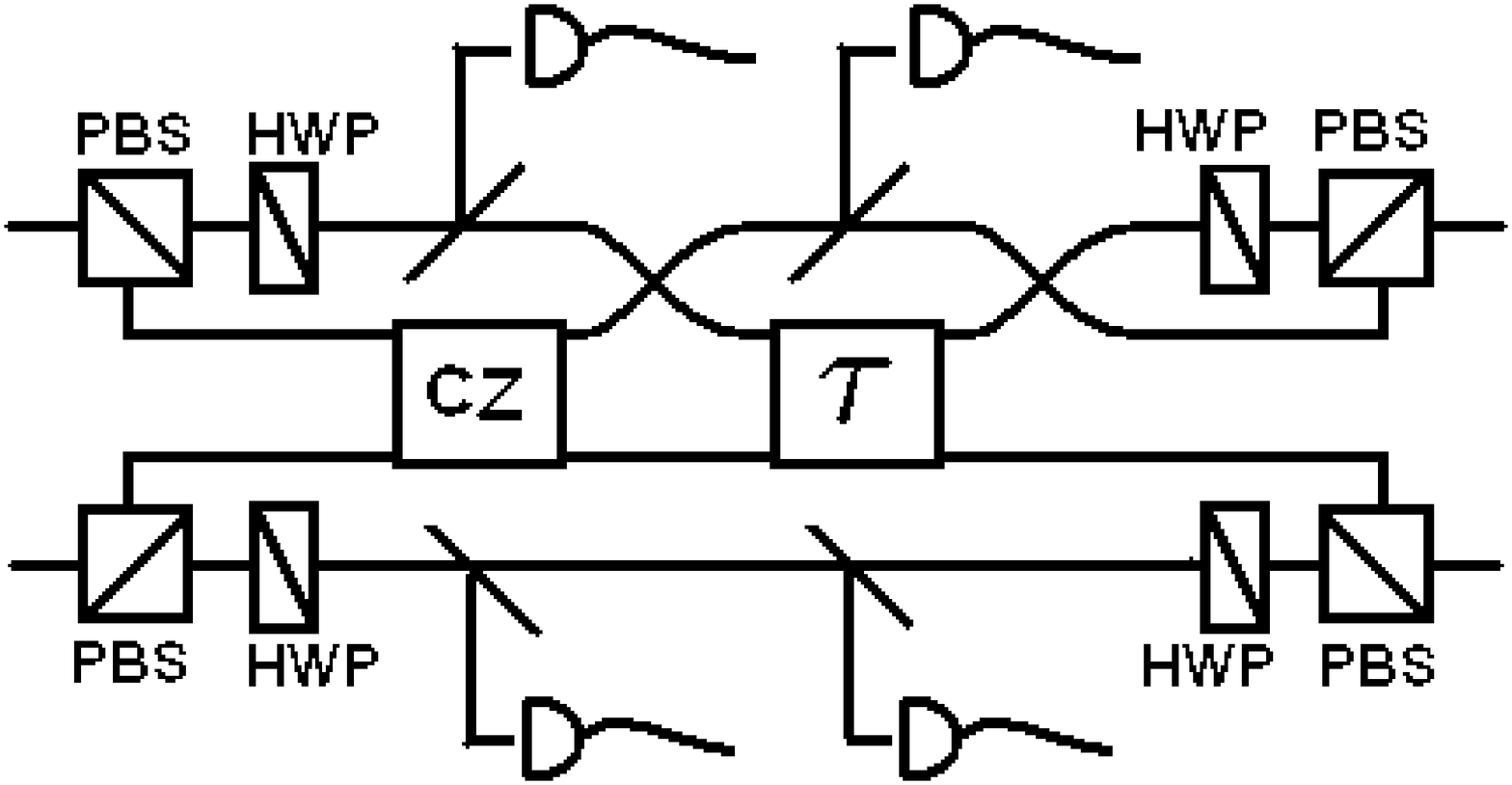,width=2cm,angle=90}}
\caption{Two-photon distillation. Schematic of operation sequence,
CZ gate, bit-flip, $\tau$-gate with two single photon distillators
and then bit-flip.} \label{fig:two_photon_distill}
\end{figure}

As shown previously, after the CZ gate and single photon
distillation, the input state
$\frac{1}{2}(|00\ra+|01\ra+|10\ra+|11\ra)$ becomes
$\frac{\gamma^{n}_{1}}{2}(|00\ra+|01\ra+|10\ra-\tau|11\ra)$. Now we
require two-photon distillation to renormalise the input state by
inducing $\tau$ on the other three computational states  as shown in
Figure~\ref{fig:two_photon_distill}. To do so, we first apply a
bit-flip on the control qubit and then apply a $\tau$-gate (see
Figure~\ref{fig:tau_gate}) and a single photon distiller on the
control qubit with transmission coefficient $\sqrt{\gamma_{1}'}$ and
another single photon distiller on the target qubit with
transmission coefficient $\sqrt{\gamma_{1}'}\tau$ and then undo the
previous bit-flip by applying another bit-flip on the control qubit.
The $\tau$-gate does the same operation as the aforementioned CZ
gate (excluding the single photon distillation) except that no minus
sign is induced on the output of $|11\bigr>$. The construction of a
$\tau$-gate is described in the next subsection. In summary the
two-photon distillation circuit does the following:

\begin{eqnarray}
|00\ra & \rightarrow & \gamma_{1}'\tau|00\ra\nonumber\\
|01\ra & \rightarrow & \gamma_{1}'\tau|01\ra\nonumber\\
|10\ra & \rightarrow & \gamma_{1}'\tau|10\ra\nonumber\\
|11\ra & \rightarrow & \gamma_{1}'|11\ra
\end{eqnarray}

\begin{figure}[h]
\centerline{\psfig{figure=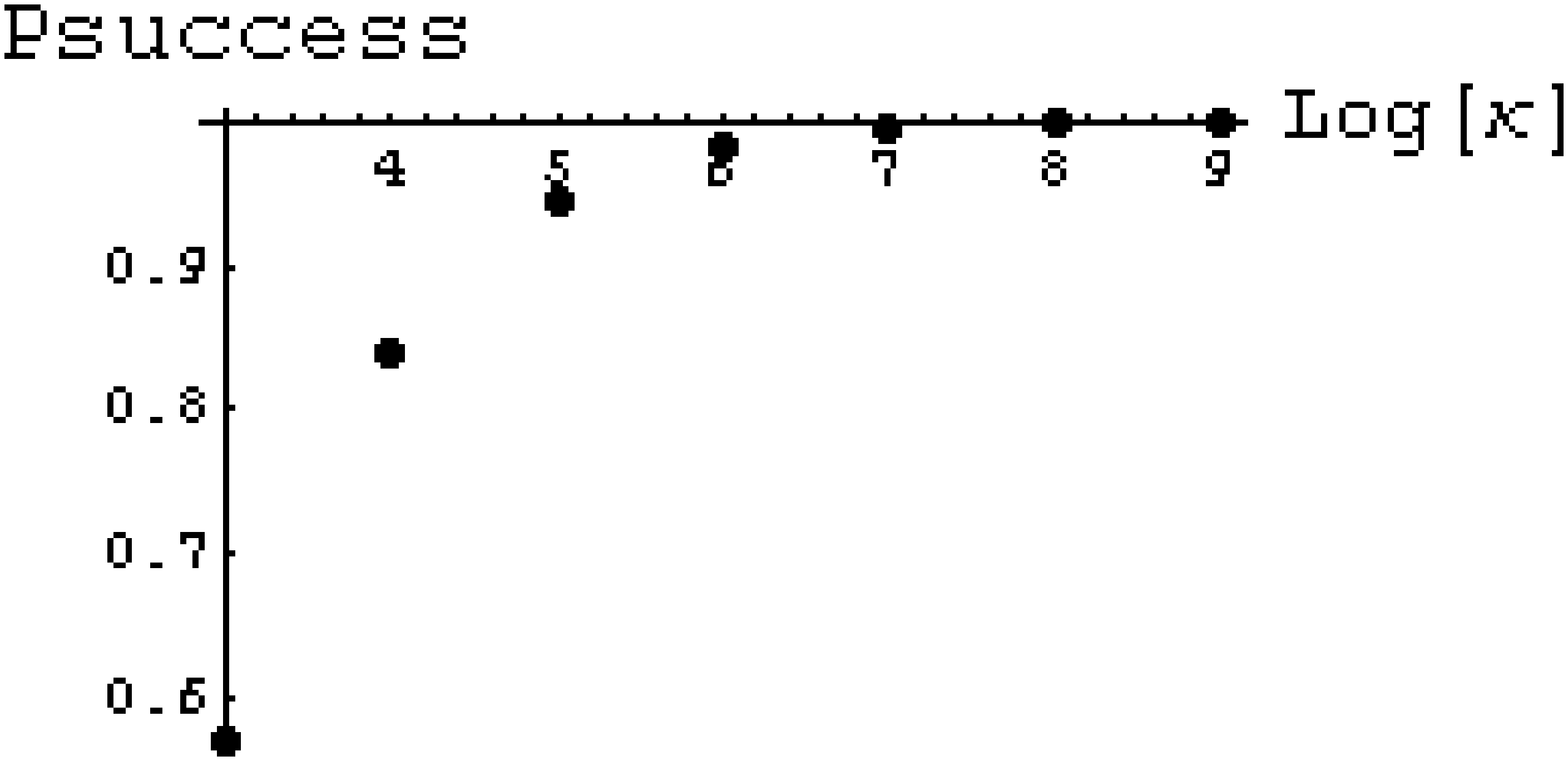,width=3cm,angle=90}}
\caption{Probability of success of the CZ gate with two-photon and
single photon distillation plotted against $\log\kappa$ (in base
10). Fidelity is always one.}\label{fig:two_photon_distill_psuccess}
\end{figure}

After the above operations, the input state
$\frac{1}{2}(|00\bigr>+|01\bigr>+|10\bigr>+|11\bigr>)$ becomes
$\frac{\gamma^{n}_{1}\gamma_{1}'\tau}{2}(|00\bigr>+|01\bigr>+|10\bigr>-|11\bigr>)$.
Now the state can be renormalised to achieve unit fidelity
\emph{independent} of $\lambda$. The explicit expression for the
probability of success is $P_s =
\gamma^{2n}_{1}\gamma_{1}'^{2}\tau^{2} =
e^{-2\lambda/\kappa}\tau^{2+2/\kappa}$
Figure~\ref{fig:two_photon_distill_psuccess} shows the probability
of success of this gate
for different values of $\kappa$.\\

\subsection{The $\tau$-Gate Circuit}

\begin{figure}[h]
\centerline{\psfig{figure=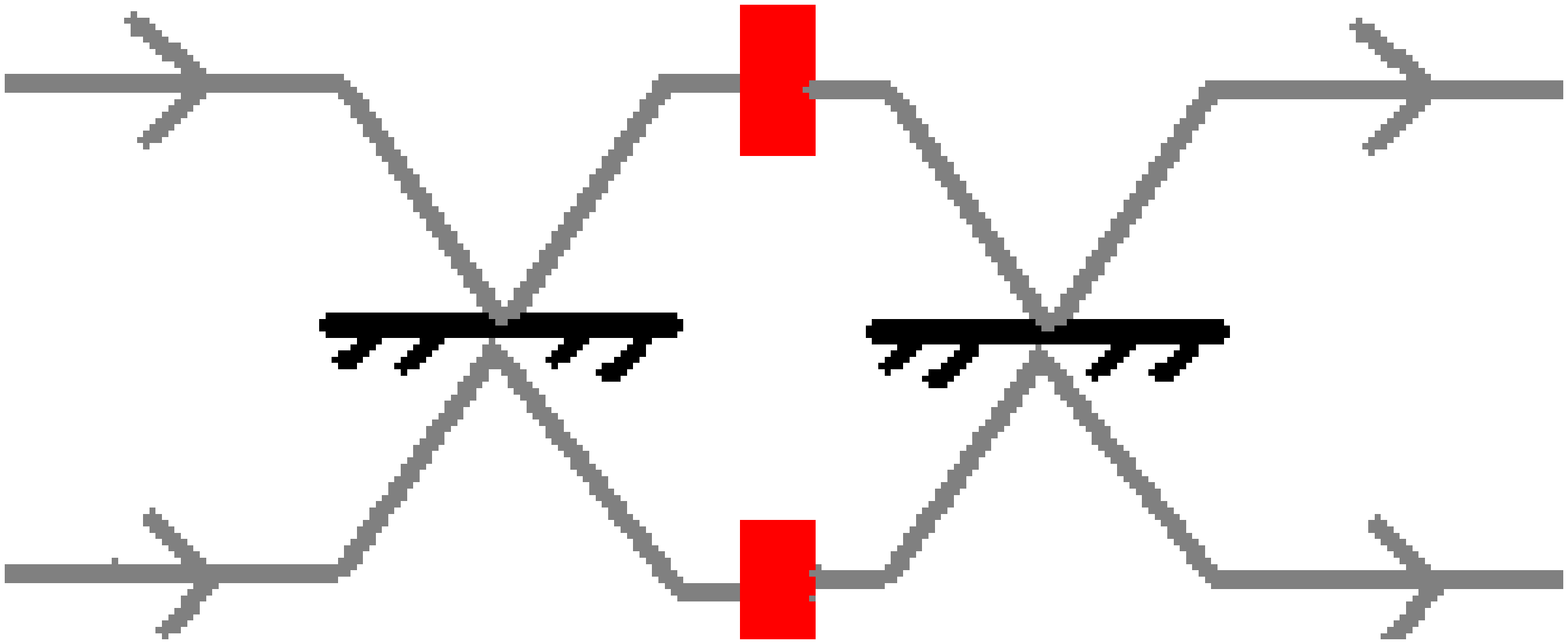,width=1cm,angle=90}}
\caption{$\tau$-gate.} \label{fig:tau_gate}
\end{figure}

\begin{figure}[h]
\centerline{\psfig{figure=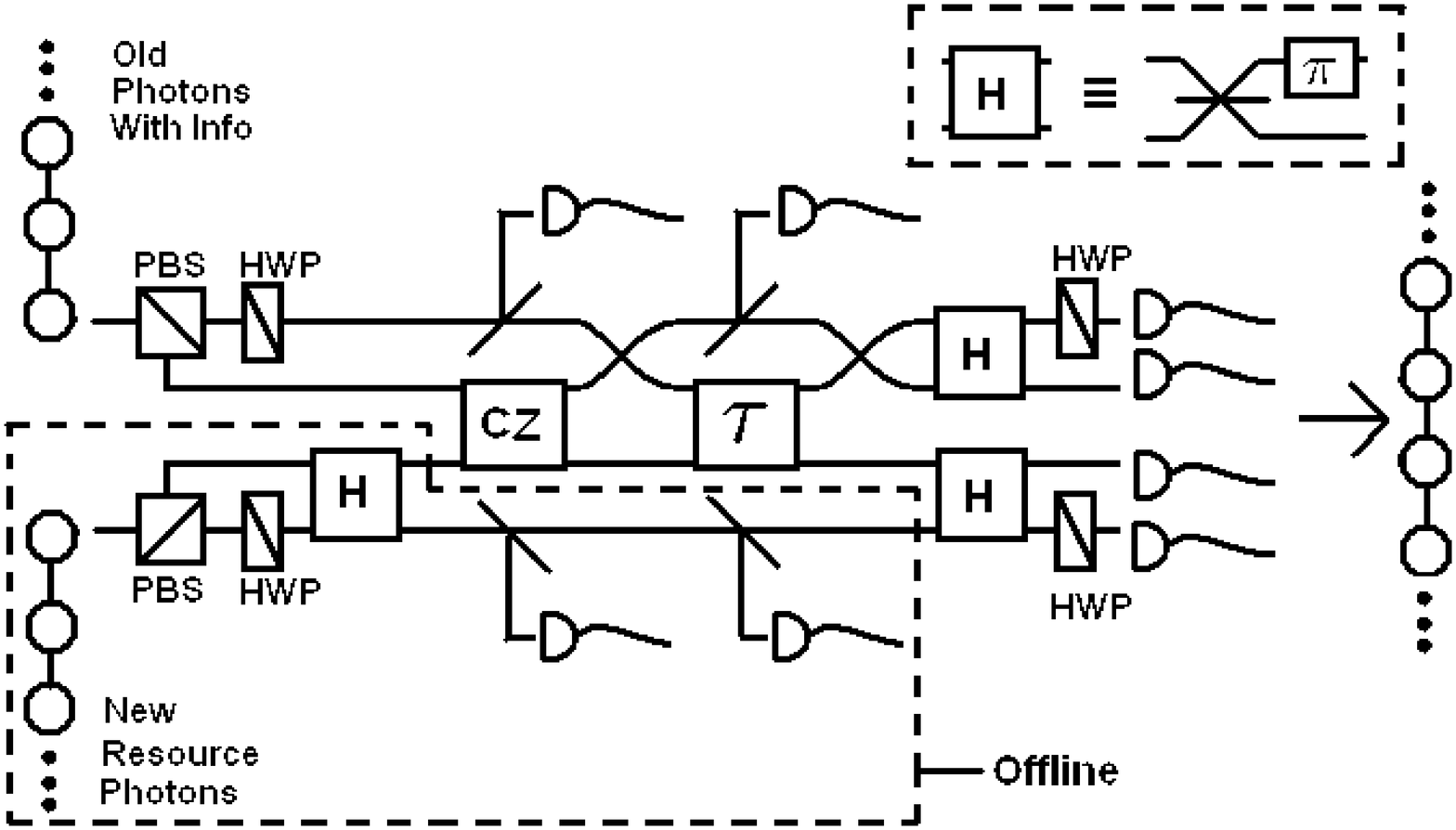,width=3.5cm,angle=90}}
\caption{Zeno Fusion gate with partial offline
distillation.}\label{fig:fusion}
\end{figure}

We can construct a $\tau$-gate with two 50-50 beam splitters, a pair
of two-photon absorbers, and some phase shifters, as shown in
Figure~\ref{fig:tau_gate}. The first beam splitter performs
$|01\ra\rightarrow|10\ra$, $|10\ra\rightarrow|01\ra$ and
$|11\ra\rightarrow\frac{i}{\sqrt{2}}(|02\ra+|20\ra)$. The pair of
two-photon absorbers then induce $\sqrt{\gamma_{1}'}$ on both
$|01\ra$ and $|10\ra$ due to single photon loss, and induce
$\gamma_{1}'\gamma_{2}'$ on $\frac{i}{\sqrt{2}}(|02\ra+|20\ra)$ due
to both single photon and two-photon loss. The second beam splitter
undoes the operation of the first beam splitter. Then with some
phase shifters to correct the relative phase between the terms and
having $\gamma_{1}'^\kappa=\gamma_{2}'=\tau$, we have a $\tau$-gate
that does the following operation:

\begin{eqnarray}
|00\ra & \rightarrow & |00\ra\nonumber\\
|01\ra & \rightarrow & \sqrt{\gamma_{1}'}|01\ra\nonumber\\
|10\ra & \rightarrow & \sqrt{\gamma_{1}'}|10\ra\nonumber\\
|11\ra & \rightarrow & \gamma_{1}'\tau|11\ra
\end{eqnarray}

\subsection{Performance of the Zeno Fusion Gate}

The fusion approach is important because it is the most efficient
method known for performing quantum computation using only linear
optics. Linear optics allows a partial Bell measurement to be made
with a probability of success of 50\% (assuming ideal detectors). In
addition the failure mode measures the qubits in the computational
basis, which does not affect the state of the remaining qubits in
the cluster or parity state. Thus a failure event only sacrifices a
single qubit from the cluster being constructed and the probability
of destroying $N$ qubits in the process of achieving a successful
fusion is $P_l = 2^{-N}$. In contrast, many of the failure events
for the Zeno gate will simply erase the photon giving no knowledge
about its state. For simplicity, and to be conservative, we will
assume all events lead to complete erasure of the photon state. In
order to recover from this situation the adjoining qubit in the
cluster must be measured in the logical basis, thus removing the
affect of the erasure \cite{ref:Lim, ref:Duan}. This means that
every failure event sacrifices two qubits from the cluster being
constructed and the probability of destroying $N$ qubits in the
process of achieving a successful fusion is $P_z = (1-P_s)^{N/2}$.
Requiring $P_l = P_z$ we estimate that the Zeno gate must have $P_s
> 0.75$ to offer an advantage over linear optics.\\

\begin{figure}[h]
\centerline{\psfig{figure=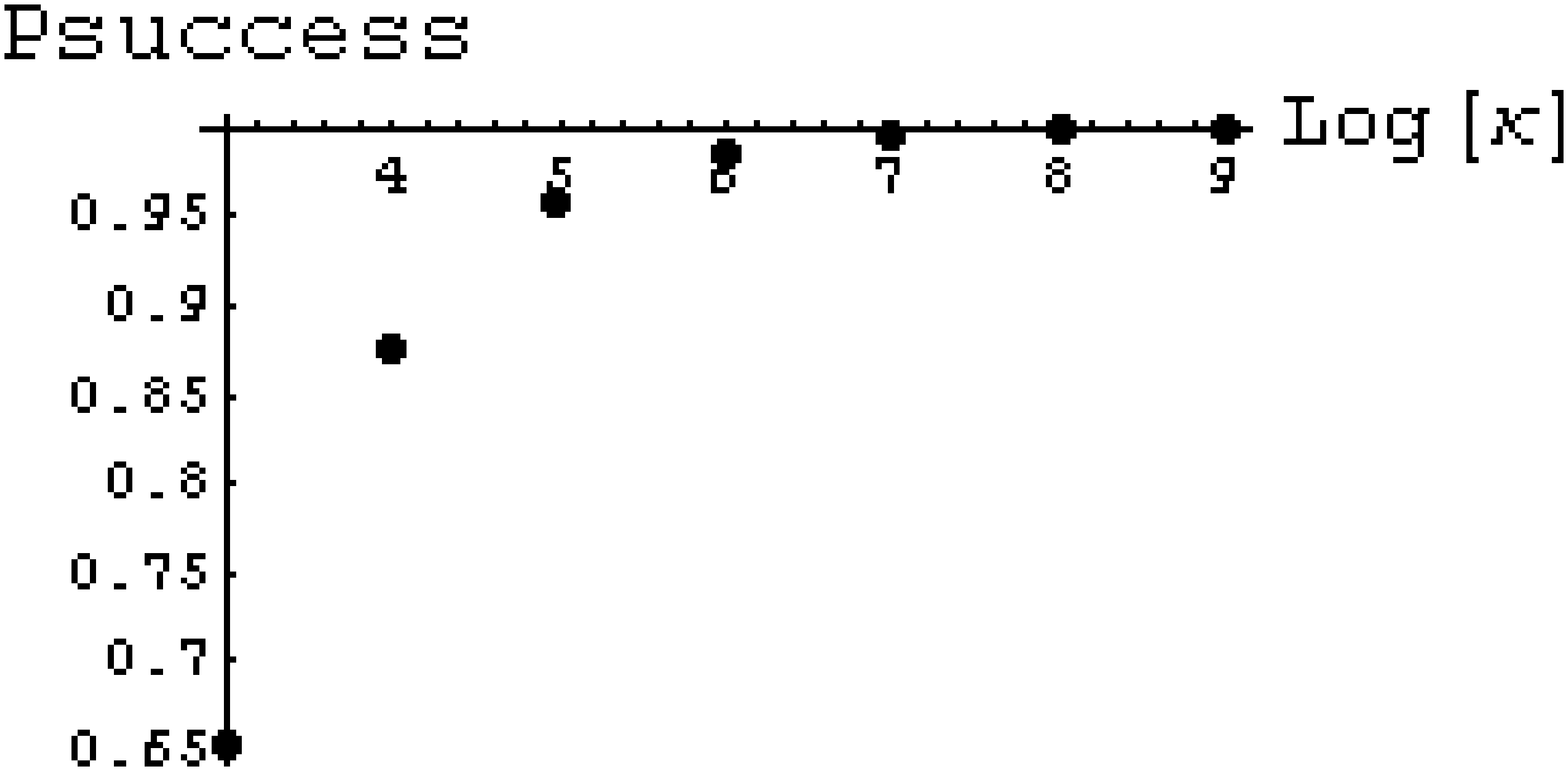,width=3cm,angle=90}}
\caption{Probability of success of the Zeno fusion gate with
partially offline two-photon and single photon distillation plotted
against $\log\kappa$ (in base 10).}\label{fig:PsuccessVsNoffline}
\end{figure}

\begin{figure}[h]
\centerline{\psfig{figure=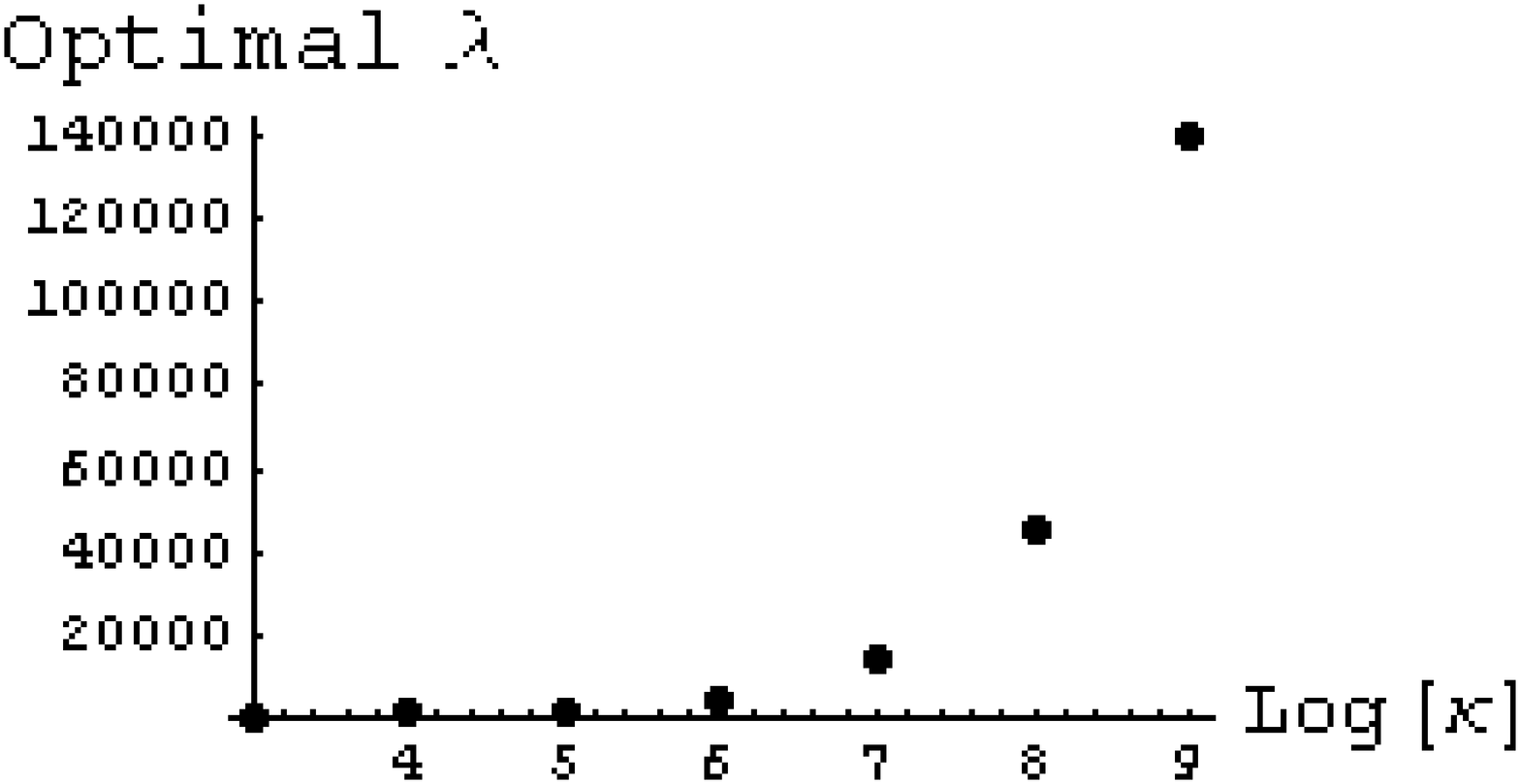,width=3cm,angle=90}}
\caption{Optimal $\lambda$ for the probability of success of the
Zeno fusion gate with partially offline two-photon and single photon
distillation plotted against $\log\kappa$ (in base
10).}\label{fig:OptimalVsNoffline}
\end{figure}

We can make one final improvement to the set-up by relocating the
distillation process for the resource qubit to offline (see Fig.5),
which boosts the probability of success. The probability of success
is then given by
$P=\frac{2\gamma_{1}^{2n}\gamma_{1}'^2\tau^2}{1+\gamma_{1}^{n}\gamma_{1}'\tau}
=
\frac{2e^{-2\lambda/\kappa}\tau^{(2+2/\kappa)}}{1+e^{-\lambda/\kappa}\tau^{(1+1/\kappa)}}$.
The plot for the probability of success versus $\kappa$ and optimal
$\lambda$ versus $\kappa$ are shown in
Figure~\ref{fig:PsuccessVsNoffline} and
Figure~\ref{fig:OptimalVsNoffline} respectively.\\

The break even point between linear optics and the Zeno gate is when
$\kappa=2200$, such that the probability of success is about $0.75$.
When $\kappa=10000$ the probability of success is about $0.87$. Thus
we conclude that an absorption ratio of ten thousand to one or more
would produce a Zeno gate with significant advantage over
linear fusion techniques.\\

\textbf{Conclusion}\\
In this paper, we have modelled Franson et al's CZ gate with a
succession of $n$ weak beam-splitters followed by two-photon
absorbers, in the (near) continuous limit of large $n$. We analysed
this CZ gate for both the ideal two-photon absorption case and the
imcomplete two-photon absorption with single photon loss case,
giving analytical and numerical results for the fidelity and
probability of success. The result shows that for a free-standing
gate we need an absorption ratio $\kappa$ of a million to one to
achieve $F>0.99$ and 100 million to one to achieve $F>0.999$, where
recent estimate only suggests that $\kappa\approx10000$ may be
achievable. We therefore employ this gate for qubit fusion, where
the requirement for $\kappa$ is less restrictive. With the help of
partially offline one-photon and two-photon distillations, we can
achieve a CZ gate with unity fidelity and with probability of
success is about 0.87 for $\kappa=10000$. We conclude that when
employed as a fusion gate, the Zeno gate could offer significant
advantages over linear techniques for reasonable parameters.\\

\textbf{Acknowledgement}\\
We thank W.J.Munro, A.Gilchrist and C.Myers for useful discussions.
This work was supported by the Australian Research Council and the
DTO-funded U.S. Army Research Office Contract No. W911NF-05-0397.\\


\begin{thebibliography}{99}
\bibitem{KNI01} E. Knill, R. Laflamme, and G.J. Milburn, \textit{Nature} \textbf{409}, 46-52 (2001).
\bibitem{OBR03} J.L.O'Brien, G.J.Pryde, A.G.White, T.C.Ralph, D.Branning, Nature {\bf 426}, 264(2003).
\bibitem{PIT03} T.B. Pittman, M.J. Fitch, B.C. Jacobs, and J.D. Franson, Phys. Rev. A {\bf 68}, 032316 (2003).
\bibitem{GAS04} S.Gasparoni, J.-W.Pan, P.Walther, T.Rudolph, and A.Zeilinger Phys. Rev. Lett. {\bf 93}, 020504 (2004).
\bibitem{YOR03} N.Yoran and B.Reznik, Phys. Rev. Lett. {\bf 91}, 037903 (2003).
\bibitem{NIE04} M.A. Nielsen, Phys. Rev. Lett. {\bf 93}, 040503 (2004).
\bibitem{HAY04} A.J.F.Hayes, A.Gilchrist, C.R.Myers and T.C.Ralph, J.Opt.B {\bf 6}, 533 (2004).
\bibitem{BRO05} D.E.Browne and T.Rudolph, Phys. Rev. Lett. {\bf 95}, 010501 (2005).
\bibitem{ref:Franson} J.D. Franson, B.C. Jacobs, and T.B. Pittman, \textit{PRA} \textbf{70}, 062302 (2004)
\bibitem{ref:Jacobs} B.C. Jacobs, T.B. Pittman, and J.D. Franson, \textit{PRA} \textbf{74}, 010303(R) (2006)
\bibitem{ref:Kaiser} W. Kaiser and C. Garrett, \textit{PRL} \textbf{7} 229 (1961)
\bibitem{ref:Thew} R.T. Thew and W.J. Munro, \textit{PRA} \textbf{63}, 030302(R) (2001)
\bibitem{LUN02} A.P. Lund, T.C. Ralph, \textit{PRA}, {\bf 66}, 032307 (2002).
\bibitem{ref:Franson3} J.D. Franson and S.M. Hendrickson \textit{quant-ph} 0603044 (2006)
\bibitem{ref:RAU01} R. Raussendorf and H.J. Briegel, \textit{PRL} \textbf{86} 5188 (2001)
\bibitem{ref:Ralph} T.C. Ralph, A.J.F. Hayes and A. Gilchrist, \textit{PRL} \textbf{95} 100501 (2005)
\bibitem{ref:Dawson} C.M. Dawson, H.L. Haselgrove, and M.A. Nielsen, \textit{PRL} 96, 020501 (2006)
\bibitem{ref:Duan} L.M. Duan and R. Raussendorf, \textit{PRL} \textbf{95}, 080503 (2005)
\bibitem{ref:Lim} Y.L. Lim, S.D. Barrett, A. Beige, P. Kok, L.C. Kwek, \textit{PRA} \textbf{73}, 012304 (2006)

\end{thebibliography}
\end{document}